\documentclass[%
 aip,
 amsmath,amssymb,
 reprint,%
]{revtex4-1}

\usepackage{graphicx}
\usepackage{dcolumn}
\usepackage{bm}

\usepackage{amsmath}
\usepackage{graphicx}
\usepackage{ dsfont }
\usepackage{dcolumn}
\usepackage{bm}
\usepackage{color,soul}
\usepackage{xcolor}
\usepackage{url}
\usepackage{tabularx}
\usepackage{array}
\usepackage{booktabs}
\usepackage{comment}
\usepackage{hyperref}
\usepackage{footnote}
\usepackage{lipsum}
\usepackage[normalem]{ulem}

\usepackage[utf8]{inputenc}
\usepackage[T1]{fontenc}
\usepackage{mathptmx}
\usepackage{etoolbox}

\makeatletter
\def\@email#1#2{%
 \endgroup
 \patchcmd{\titleblock@produce}
  {\frontmatter@RRAPformat}
  {\frontmatter@RRAPformat{\produce@RRAP{*#1\href{mailto:#2}{#2}}}\frontmatter@RRAPformat}
  {}{}
}%
\makeatother
\begin{document}

\preprint{AIP/123-QED}

\title{Homotopy transitions and 3D magnetic solitons}
\author{V. M. Kuchkin}%
\email{v.kuchkin@fz-juelich.de}
 \affiliation{Peter Gr\"unberg Institute and Institute for Advanced Simulation, Forschungszentrum J\"ulich and JARA, 52425 J\"ulich, Germany}
 \affiliation{Department of Physics, RWTH Aachen University, 52056 Aachen, Germany}
\author{N. S. Kiselev}%
 \affiliation{Peter Gr\"unberg Institute and Institute for Advanced Simulation, Forschungszentrum J\"ulich and JARA, 52425 J\"ulich, Germany}

\date{\today}

\begin{abstract}

This work provides a concept for three-dimensional magnetic solitons based on mapping the homotopy path between various two-dimensional solutions onto the third spatial axis. 
The representative examples of statically stable configurations of that type in the model of an isotropic chiral magnet are provided. Various static and dynamic properties of such three-dimensional magnetic solitons are discussed in detail.


\end{abstract}

\maketitle


\section{\label{sec:intro}Introduction}

Chiral magnets are a distinct class of materials, where the competition between the Heisenberg exchange and the chiral Dzyaloshinskii-Moria interaction~\cite{Dzyaloshinskii, Moriya} (DMI) gives rise to a wide diversity of magnetic solitons -- localized magnetic textures with particle-like properties.
The most representative example of magnetic solitons in chiral magnets are magnetic skyrmions which have been intensively studied both theoretically~\cite{Bogdanov89,Bogdanov1994,Bogdanov1999} and experimentally~\cite{Muhlbauer2009,Tonomura2012,Lebech1989,Wilhelm2011,Yu2011,Yu2010,Onose2012} in the last decades. 
In thick films and bulky samples, the magnetization vector field of chiral skyrmions resembles a filamentary structure composed of vortex-like tubes.

More recent studies have revealed a diversity of 3D solitons in chiral magnets beyond skyrmions.
The most prominent examples are chiral bobbers~\cite{Rybakov2015, Zheng18}, skyrmion bags~\cite{Rybakov2019, Foster_2019, Kuchkin2020_2, tang21}, heliknotons~\cite{Smalykh_2020}, and skyrmion braids~\cite{zheng21}.
Because of distinct topology, the above solitons possess different static, dynamic, and transport properties. 
In particular, the chiral bobbers are distinct by the presence of magnetic singularity -- Bloch point, while other magnetic textures are characterized by smooth magnetic vector fields.
Because of that, contrary to other solitons, the chiral bobber is not a topological soliton.
The skyrmions and heliknotons, on the other hand, belong to different topological groups characterized by distinct topological invariants. 

In this work, we present another type of magnetic solitons which belong to a topological group of skyrmions and are stabilized in bulk crystals of chiral magnets.
Because of the unique properties of these solutions, it is reasonable to attribute them to a distinct class of magnetic solitons.
Here we will refer to them as \textit{hybrid} skyrmion tubes.
The crosssections of well-studied skyrmion tubes usually represent nearly identical configurations with small additional modulations due to the braiding effect~\cite{zheng21} or the presence of the noncollinear background~\cite{Du18} and/or free surfaces~\cite{Rybakov13}.
We show that besides such nearly homogeneous tubes, there are also solutions where the crosssections represent a continuous transformation between the skyrmions of different configurations but the same topological index -- skyrmions of the same homotopy group.
In a representative example, we illustrate the stability of such exotic spin textures and discuss their unique dynamic properties.
Besides that, we discuss an important consequence of applying such a homotopical concept to truly 3D localized magnetic spin textures. 
In particular, we present the solution representing a topologically trivial but statically stable and truly three-dimensional soliton -- 3D chiral droplet.

\section{\label{sec:model}Model}

The model Hamiltonian for chiral magnet with bulk type DMI has the folowing form:
\begin{equation}
E = \displaystyle\int_{V_\mathrm{m}}(\mathcal{A}(\nabla {\bf n})^2+\mathcal{D}{\bf n}\cdot \nabla {\bf n} + U({\bf n}))\mathrm{d}V_\mathrm{m},\label{E_micro}
\end{equation}
where ${\bf n}={\bf M}/M_\mathrm{s}$ is normalized magnetization $|{\bf n}|=1$, $\mathcal{A}$ and $\mathcal{D}$ are micromagnetic constants of the Heisenberg exchange and the DMI, respectively.
$V_\mathrm{m}$ is the volume of the magnetic sample.
The potential energy term $U({\bf n})$ includes an uniaxial (easy-axis or easy-plane) magnetic anisotropy, $\mathcal{K}_\mathrm{u}$, and the interaction with the external magnetic field, $\mathbf{B}_\mathrm{ext}\parallel{\bf e}_\mathrm{z}$:
\begin{equation}
 U({\bf n}) = -\mathcal{K}_\mathrm{u}n_\mathrm{z}^2 - M_\mathrm{s}B_\mathrm{ext}n_\mathrm{z}.\label{potential_term}
\end{equation}

Following the standard procedure~\cite{Bogdanov89, Bogdanov1994, Bogdanov1999}, the functional \eqref{E_micro} can be written in a more suited for analysis, dimensionless form
\begin{equation}
\mathcal{E} =\!\displaystyle\int_{V}\!\left(\dfrac{(\nabla {\bf n})^{2}}{2}+2\pi{\bf n}\cdot \nabla {\bf n} -4\pi^{2}(u n_\mathrm{z}^2 + h n_\mathrm{z}) \right)\mathrm{d}V,\label{EE0}
\end{equation}
where $\mathcal{E}=E/2\mathcal{A}$ is the reduced energy,   $u=\mathcal{K}_\mathrm{u}/M_\mathrm{s}B_\mathrm{D}$ and $h=B_\mathrm{ext}/B_\mathrm{D}$ are the dimensionless values
of anisotropy constant and the strength of the external magnetic field, respectively.
The integration in \eqref{EE0} is over the reduced volume
$V=V_\mathrm{m}/L_\mathrm{D}^3$.
The characteristic parameters $L_\mathrm{D}=4\pi\mathcal{A}/\mathcal{D}$ and  $B_\mathrm{D}=\mathcal{D}^2/2 M_\mathrm{s}\mathcal{A}$ are the period of spin spiral and the saturation field for isotropic case, $\mathcal{K}_\mathrm{u}=0$.

The energy functional \eqref{EE0} remains valid for the 2D (or quasi 2D) systems where the magnetization does not change along the $z$-axis and integration is carried out over $\mathrm{d}V = l\mathrm{d}x\mathrm{d}y$, where $l$ is the film thickness. 
The 2D model of the chiral magnet, besides ordinary axially symmetric $\pi$-skyrmions, provides a large variety of magnetic solitons \textit{e.g.} skyrmion bags~\cite{Rybakov2019, Foster_2019} and skyrmions with chiral kinks~\cite{Kuchkin2020_2}.
The homotopy classification of localized solutions in 2D is provided by the following invariant
\begin{equation}
Q = \frac{1}{4\pi}\int  \mathbf{n}\cdot \left(\partial_x\mathbf{n}\times\partial_y\mathbf{n}\right)\,  \mathrm{d}x\mathrm{d}y.
\label{Q}
\end{equation}
The solutions with identical integer index $Q$ are called the solutions of one homotopy class.
The latter implies that one can continuously transform the vector fields of such solutions into each other. 
For instance, let us assume there are two localized magnetic textures ${\bf n}_{1}(x,y)$ and ${\bf n}_{2}(x,y)$ of identical charge $Q$.
Then, there is a vector field transformation, ${\bf n}(x,y;s)$, where $s\in[0,1]$, such that for ${\bf n}(x,y;0)={\bf n}_{1}(x,y)$ and ${\bf n}(x,y;1)={\bf n}_{2}(x,y)$. When ${\bf n}(x,y;s)$ is differentiable with respect to $x$, $y$, and $s$ at any point such transformation can be called a homotopy path.
Since for any ${\bf n}_{1}(x,y)$ and ${\bf n}_{2}(x,y)$, there is an infinite number of homotopy paths, there is no unique method to construct such paths.
To make the search more definite, we consider only those homotopy paths that also satisfy the minimum energy path (MEP) criterion.
In particular, we use geodesic nudged elastic band (GNEB) method~\cite{Bessarab_2015,Bessarab_2017} implemented in Spirit code~\cite{Spirit}. 
The details of the MEP calculation are provided in the Appendix \ref{A_MEP}.

\begin{figure}[ht!]
\includegraphics[width=8cm]{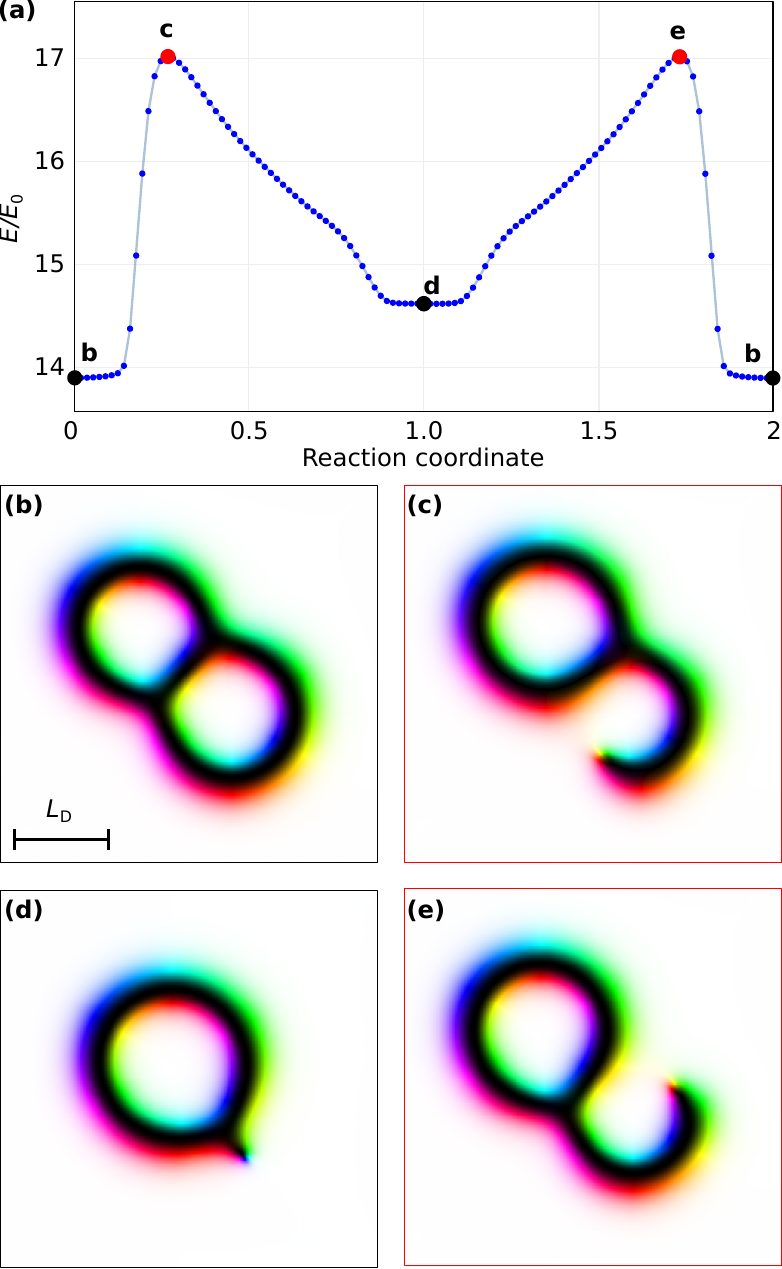}
\caption{\textbf{a} MEP between two stable skyrmion states with $Q=1$ depicted in \textbf{b} and \textbf{d}. The spin configuration corresponding to the saddle points are depicted in \textbf{c} and \textbf{e}.
The reaction coordinate is given in reduced unit with respect to its value at intermediate state depicted in \textbf{d}.
The calculations are performed at $h=0.627$ and $u=0$. The energy in \textbf{a} is given with respect to the energy of ferromagnetic state, $E_0$.
\label{Fig1}
}
\end{figure}

\section{Results}

In Fig.~\ref{Fig1}\textbf{a} we provide representative example of the MEP between two skyrmions with $Q=1$. We show only the spin texture corresponding to local minima and saddle points. 
Noticeably, the textures corresponding to the central minimum state (\textbf{d}) and saddle points \textbf{(c)}, \textbf{(e)} contain a chiral kink~\cite{Kuchkin2020_2} while another minimum state is a skyrmion bag free of kinks \textbf{(b)}.
The MEP presented in Fig.~\ref{Fig1}\textbf{a} satisfy the above criteria for the homotopy path. The parameter $s$ can be associated with the reaction coordinate which has a meaning of the relative distance between the images (snapshots of the vector field) in the multidimensional parameter space.

\begin{figure*}[ht!]
\includegraphics[width=17cm]{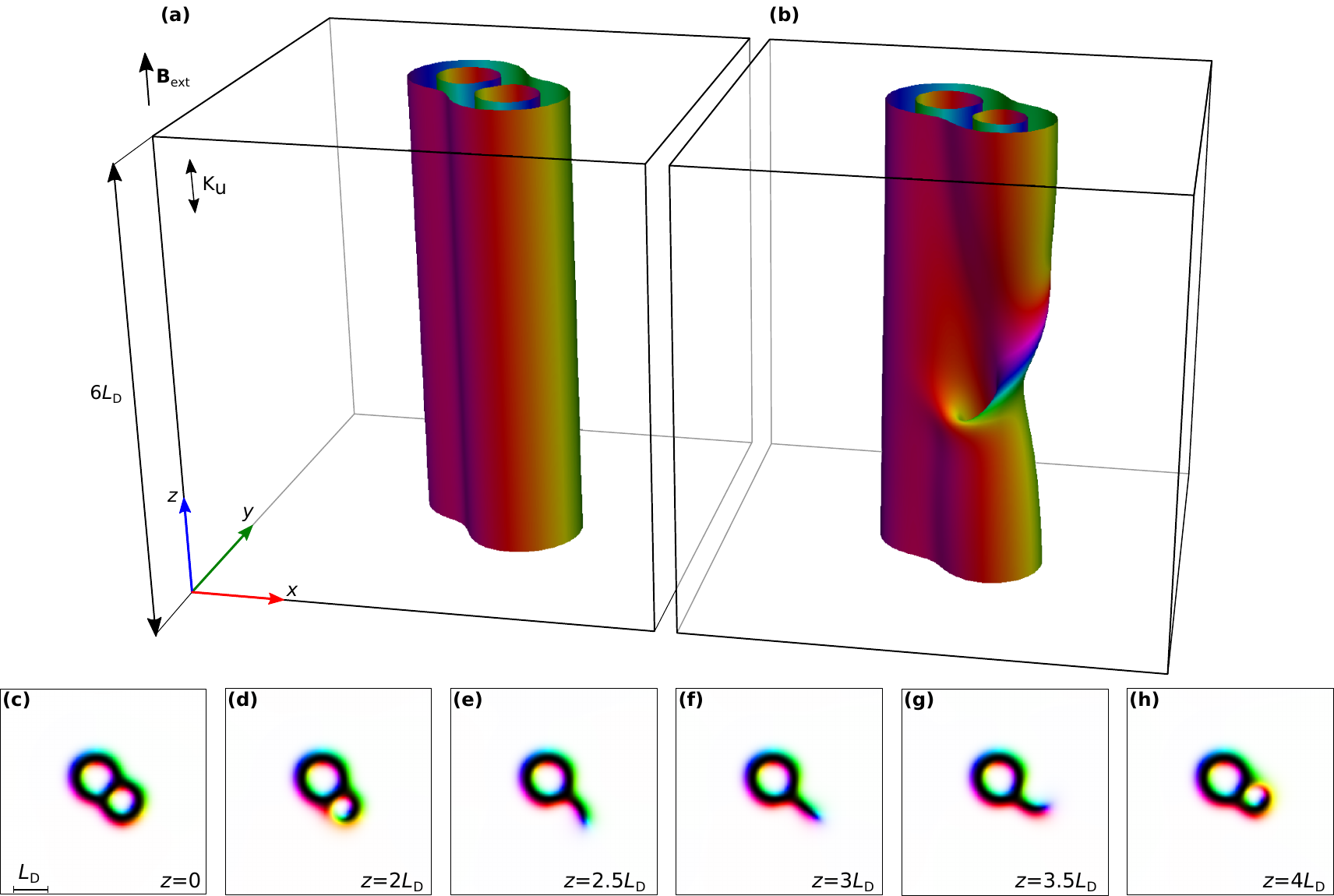}
\caption{\label{fig:1}\textbf{a} and \textbf{b} show isosurfaces $n_\mathrm{z}=0$ for ordinary and hybrid skyrmion tubes stabilized at $h=0.45$, $u=0.3$ in the box of size $L_\mathrm{x}=L_\mathrm{y}=L_\mathrm{z}=6L_\mathrm{D}$, $L_\mathrm{D}=32a$ with periodic boundary conditions in all directions.
\textbf{b}-\textbf{g} show the cross-sections of the hybrid tube, while the ordinary skyrmion tube is characterized by identical magnetic texture (alike \textbf{c}) in every $z$ section.}
\end{figure*}

To construct an initial configuration for the 3D skyrmion tube, one can take the stable 2D skyrmion configuration and place it at each $xy$-plane of the 3D simulated domain.
Statically stable configuration of such homogeneous skyrmion tube corresponding to the 2D skyrmion in Fig.~\ref{Fig1}\textbf{(b)} is provided in Fig.~\ref{fig:1}\textbf{(a)}.
To construct the initial state for a nonhomogeneous 3D skyrmion tube, we use a mapping from the homotopy path to the third spatial axis, $s\rightarrow z$. In other words, to create a 3D magnetic texture, we sequentially lay down the spin texture of the images from MEP on top of each other along with the $z$-axis.
The statically stable spin configuration obtained by the energy minimization of that initial state is provided in Fig.~\ref{fig:1}\textbf{(b)} and its crosssections are shown in \textbf{(c)}-\textbf{(h)}.
The intermediate region resembling a knot on the isosurface of the skyrmion string in Fig.~\ref{fig:1}\textbf{(b)} is well localized and thus can be thought of as a soliton that is hosted by a skyrmion string.
This nonhomogeneous 3D skyrmion tube is a representative example of the solutions to which we refer as \textit{hybrid} skyrmion tubes.
For conciseness, the localized intermediate region, we call a \textit{knot} in the following.
The choice of such terminology is justified by pure visual analogy and has nothing to do with the topological knots.
Whether one can continuously unwind that knot without the appearance of Bloch points represents an intriguing question that, however, goes out of the scope of the present work.

At the strong magnetic field,  $h\geq1-2u$, the hybrid skyrmion tubes represent a metastable state embedded in the ferromagnetic vacuum.
For $h<1-2u$, the vacuum for such states is the cone phase.
The range of fields and anisotropies where hybrid skyrmion tubes remain stable depends on the particular configuration -- the type of 2D skyrmions in the tube crosssections.
For instance, the hybrid skyrmion tube depicted in Fig.~\ref{fig:1}\textbf{(b)} at $u=0$ remain stable, at least in the range of $0<h<0.35$.

In general, the hybrid skyrmion tubes can be composed of a few intermediate states representing stable 2D skyrmions.
Here we consider the solution composed of two intermediate states only.
Such configurations are easier to handle because they satisfy periodical boundary conditions.
It is worth noting, the stabilization of hybrid skyrmion tubes does not require the periodic boundary conditions along $z$-axis.
The latter makes the experimental observation of hybrid skyrmion tubes in thick films of chiral magnets quite promising.
In the case of free boundaries along $z$-axis, the tube has an additional surface twist modulations.
For thick enough films ($\sim 6L_\mathrm{D} = 420$ nm for FeGe) 
These modulations do not decrease the tube stability.
On the other hand the presence of the free boundaries the hybrid skyrmion tube depicted in Fig.~\ref{fig:1} \textbf{(b)} can continuously unwind into the host skyrmion tube shown in Fig.~\ref{fig:1} \textbf{(a)}. 
%
This process is illustrated by Supplementary Movie 1, where each frame was obtained by a manual shift of the entire spin texture along the $z$-axis and following incomplete relaxation of the system. In the case of complete relaxation, the system converges to its initial state, which indicates the presence of the energy barrier for the knot to escape through the free boundary.

To demonstrate the unique dynamic properties of hybrid skyrmion tubes we perform micromagnetic simulations based on the Landau-Lifshitz-Gilbert equation~\cite{Landau1965}:
\begin{equation}
  \dfrac{\partial\mathbf{n}}{\partial t}=-\gamma\mathbf{n}\times\mathbf{H}_\mathrm{eff}+\alpha\mathbf{n}\times\dfrac{\partial\mathbf{n}}{\partial t} -\textbf{T}_\mathrm{ZL},\label{LLG}
\end{equation}
where $\gamma$ is the gyromagnetic ratio, $\alpha$ is the Gilbert damping,    $\mathbf{H}_\mathrm{eff}=-\dfrac{1}{M_\mathrm{s}}\dfrac{\delta\mathcal{E}}{\delta\mathbf{n}}$ is the effective field.
The last term in \eqref{LLG} is the Zhang-Li torque~\cite{ZhangLi2004} because of the electric current:
\begin{equation}
\textbf{T}_\mathrm{ZL}= \mathbf{n}\!\times\!\left[\mathbf{n}\times\left(\mathbf{I}\cdot\nabla\right)\mathbf{n}\right]+\xi\,\mathbf{n}\times\left(\mathbf{I}\cdot\nabla\right)\mathbf{n},
\label{T_ZL}
\end{equation} 
where the vector $\mathbf{I}=\mathbf{j}{\mu_\mathrm{B}p}(1\!+\!\xi^{2})^{-1}(e M_\mathrm{s})^{-1}$ is proportional to the current density $\mathbf{j}$,
$\xi$ is the degree of non-adiabaticity,  $p$ is the polarization of the spin current, $\mu_\mathrm{B}$ is the Bohr magneton and  $e$ is the electron charge. 

When the 3D soliton is embedded in the FM state, the current direction can be chosen arbitrary. 
On the contrary, when the soliton is embedded in non homogeneous vacuum, e.g., cone phase, it is essential to chose the current direction perpendicular to the $\mathbf{q}$-vector of the cone to prevent the excitation of the background.
To simplify the following discussions we consider the case of hybrid skyrmion tube in ferromagnetic vacuum. 

Supplementary Movie 2 illustrates the dynamics of the hybrid skyrmion tube when the current is along the $z$-axis. 
The simulations were performed with Mumax~\cite{Mumax} code.
The example of Mumax script and the initial state files are provided in Supplementary Data.
We used periodical boundary conditions in all spatial directions to simulate a bulk crystal.
Under the electric current parallel to the skyrmion tube, the knot on the isosurfaces moves along the tube in the direction opposite to the current.
In this case, the position of the host skyrmion tube in the $xy$-plane remains fixed.

In Supplementary Movie 3, we show the dynamics of the skyrmion tube when the current is perpendicular to the skyrmion tube, $\mathbf{I}\parallel\mathbf{e}_\mathrm{x}$.
In this case, both the host skyrmion tube and the knot move.
As a result, all three components of the knot velocity are non-zero. 

To quantify the knot velocity, we follow the approach used in our previous work on 2D skyrmion dynamics~\cite{Kuchkin_2021}. 
Extending this approach to 3D textures the position of the knot $\mathbf{R}=\left(R_\mathrm{x}, R_\mathrm{y}, R_\mathrm{z}\right)$ can be defined as follows
\begin{equation}
    R_k = \dfrac{L_k}{2\pi}\tan^{-1}\dfrac{\int\mathcal{N}_{ij}(r_{k})\sin\left(2\pi r_k/L_{k}\right)\mathrm{d}r_{k}}{\int\mathcal{N}_{ij}(r_{k})\cos\left(2\pi r_k/L_{k}\right)\mathrm{d}r_{k}}\pm l_{k} L_{k},
\end{equation}
where $\mathcal{N}_{ij}(r_{k})=\int n_\mathrm{z}(\mathbf{r})\mathrm{d}r_{i}\mathrm{d}r_{j}$ is magnon density averaged in $r_{k}$-plane, indices $i, j, k \in \{\mathrm{x},\mathrm{y},\mathrm{z}\}$.
Sigh $\pm$ accounts for the direction of solitons motion: along the basis vector $\mathbf{e}_k$ $(+)$ or in the opposite direction $(-)$.
The integer $l_k$ is the number of times the soliton crossed the corresponding domain boundary.
By tracing the position $\mathbf{R}$ at each moment in time, we can estimate the instant velocity of the knot, $\mathbf{v}=d\mathbf{R}/dt$, and compare it with the results of a semi-analytical approach based on the method of collective coordinates suggested by Thiele~\cite{Thiele1973}. 
Assuming the rigid motion of magnetic textures with velocity $\mathbf{v}$, i.e. $\mathbf{n}(\mathbf{r},t)=\mathbf{n}(\mathbf{r}-\mathbf{v}t)$, from \eqref{LLG} and \eqref{T_ZL}, one can derive the Thiele equation:
\begin{equation}
\mathbf{G}\times(\mathbf{v}+\mathbf{I})+\hat{\Gamma}(\alpha\mathbf{v}+\xi\mathbf{I})=0,
\label{Thiele_eq}
\end{equation} 
where the gyro-vector, $\mathbf{G}$, and the dissipation tensor, $\hat{\Gamma}$, have components $G_{i}$ and $\Gamma_{ij}$ defined as:
\begin{gather}
G_{i}=-\displaystyle\int\epsilon_{ijk}\epsilon_{i^{\prime}j^{\prime}k^{\prime}}n_{i^{\prime}}\dfrac{\partial n_{j^\prime}}{\partial r_j}\dfrac{\partial n_{k^\prime}}{\partial r_k}\mathrm{d}V,\label{G}\\
\Gamma_{ij}=\displaystyle\int\dfrac{\partial n_{k}}{\partial r_i}\dfrac{\partial n_{k}}{\partial r_j}\mathrm{d}V,
\label{DT}
\end{gather} 
where $\epsilon_{ijk}$ is Levi-Civita symbol.
The advantage of the Thiele approach is that the solution for the soliton velocity can be written explicitly. For instance, for $\mathbf{I}=I\mathbf{e}_\mathrm{x}$, the solution of \eqref{Thiele_eq} for the uniform skyrmion tubes, $\Gamma_\mathrm{xz} = \Gamma_\mathrm{yz} = \Gamma_\mathrm{zz} = 0$ takes a form identical to the 2D case 
\begin{align}
& v_\mathrm{x}=-I\dfrac{G_\mathrm{z}^{2} + G_\mathrm{z}\Gamma_\mathrm{xy}(\xi\!-\!\alpha) + \alpha\xi\mathrm{det}\hat{\Gamma}}{G_\mathrm{z}^{2} + \alpha^{2}\mathrm{det}\hat{\Gamma}},\nonumber\\
& v_\mathrm{y}=-I\dfrac{G_\mathrm{z}\Gamma_\mathrm{xx}(\xi\!-\!\alpha)}{G_\mathrm{z}^{2} + \alpha^{2}\mathrm{det}\hat{\Gamma}},\label{sol_2D}\\
& v_\mathrm{z}=0.\nonumber
\end{align}
For the habridized skyrmion tube all the entries of the dissipation tensor $\hat{\Gamma}$ have non-zero values and the solution of \eqref{Thiele_eq} takes the form
\begin{align}
& v_\mathrm{x}=-I\dfrac{G_\mathrm{z}\Gamma_\mathrm{xx}\!+\!(\Gamma_\mathrm{xy}\Gamma_{zz}\!-\!\Gamma_\mathrm{xz}\Gamma_\mathrm{yz})(\xi\!-\!\alpha)\!+\!\alpha\xi\mathrm{det}\hat{\Gamma}}{G_\mathrm{z}^{2}\Gamma_\mathrm{zz}+\alpha^{2}\mathrm{det}\hat{\Gamma}},\nonumber\\
& v_\mathrm{y}=-I\dfrac{ G_\mathrm{z}(\Gamma_\mathrm{xz}^{2}-\Gamma_\mathrm{xx}\Gamma_\mathrm{zz})(\xi\!-\!\alpha)}{G_\mathrm{z}^{2}\Gamma_\mathrm{zz}+\alpha^{2}\mathrm{det}\hat{\Gamma}},\label{sol_3D}\\
& v_\mathrm{z}=-I\dfrac{ G_\mathrm{z}(G_\mathrm{z}\Gamma_\mathrm{xz}/\alpha + \Gamma_\mathrm{xx}\Gamma_\mathrm{yz}-\Gamma_\mathrm{xz}\Gamma_\mathrm{xy})(\xi\!-\!\alpha)}{G_\mathrm{z}^{2}\Gamma_\mathrm{zz}+\alpha^{2}\mathrm{det}\hat{\Gamma}}.\nonumber
\end{align}

\begin{figure}[ht!]
\includegraphics[width=7.7cm]{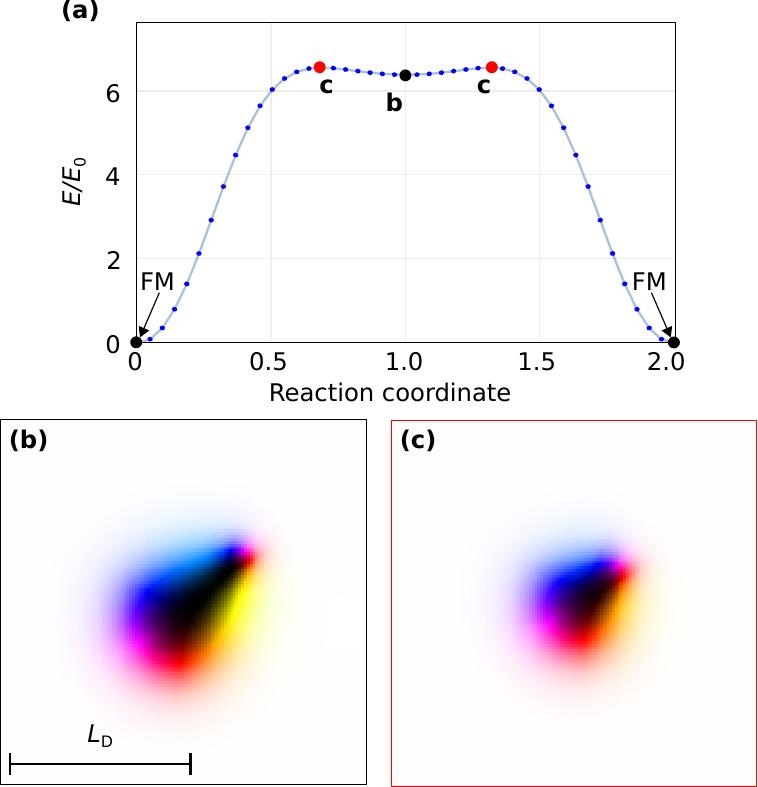}
\caption{\textbf{a} MEP between chiral droplet with $Q=0$ depicted in \textbf{b} and FM state.
The spin configuration corresponding to the saddle point is depicted in \textbf{c}.
The reaction coordinate is given in reduced unit with respect to its value at intermediate state depicted in \textbf{b}.
The calculations are performed at $h=0.65$ and $u=0$.
The energy in \textbf{a} is given with respect to the energy of ferromagnetic state, $E_0$.
\label{Fig3}}
\end{figure}

It is worth noting two remarkable features of the solution \eqref{sol_3D}.
Firstly, there is no continuous transition between \eqref{sol_3D} and \eqref{sol_2D} even when $\Gamma_\mathrm{xz}$, $\Gamma_\mathrm{yz}$ $\Gamma_\mathrm{zz}$ tend to zero.
Secondly, in the most general case of 3D soliton, the components of gyro-vector $\mathbf{G}$ can be thought of as 2D topological charges in corresponding $x$, $y$ and $z$ crosssections.
For hybrid skyrmion tubes, however, similar to the 2D case, only the third component of the gyro-vector is non zero and proportional to 2D topological charge in $xy$-plane, $\mathbf{G}=4\pi Q L_\mathrm{z}\mathbf{e}_\mathrm{z}$.
%

To calculate the velocity components in \eqref{sol_3D}, one has to find the components of the gyro-vector~\eqref{G} and the dissipation tensor~\eqref{DT} for a particular magnetization distribution.
For that, we use the magnetization vector field at the static equilibrium obtained by the numerical energy minimization of the functional \eqref{E_micro}. 
It is worth emphasizing that the soliton velocity obtained from the solutions of the Thiele equation \eqref{Thiele_eq} corresponds to the velocities in a steady-state only. Thus the velocities estimated from LLG simulations can be compared with the analytical solutions only when the transient state is over, and all components of the $\mathbf{v}$ reached the saturation.

In the most general case of 3D soliton, when all three components of the velocity are non-zero, one can introduce two deflection angles $\beta_{1}=\arctan\left(v_\mathrm{y}/v_\mathrm{x}\right)$ and $\beta_{2}=\pi/2 - \arctan\left(\sqrt{v_\mathrm{x}^2 +v_\mathrm{y}^2 }/v_\mathrm{z}\right)$.
The angle $\beta_{1}$ is also known as skyrmion Hall angle.
Because of that we refer to $\beta_{2}$ as the second skyrmion Hall angle.
Table~\ref{tab:skyrmion_dyn} demonstrates a good agreement between the values of $\beta_1$ and $\beta_2$ estimated from LLG simulations and calculated with the Thiele approach for parameters $\mathbf{j}=-5\cdot10^{8}\mathbf{e}_\mathrm{x}$ A/m$^2$, $\alpha=0.01$ and $\xi=0.05$. 
Noticeably, for particular tube depicted in Fig.~\ref{fig:1}(b), the angle $\beta_{2}$ is larger than $\beta_{1}$.
As follows from the Thiele equation \eqref{Thiele_eq}, when $\mathbf{I}=I\mathbf{e}_\mathrm{z}$, in agreement with the micromagnetic simulations, only the $z$-component of the knot velocity is non zero, $v_\mathrm{z}=-I\xi/\alpha$.
\begin{table}[ht]
	\caption{Two Hall angles for hybrid skyrmion tube estimated by the LLG simulations and with the Thiele method of collective coordinates}
	\label{tab:skyrmion_dyn}
	\centering
	\begin{tabular}{| c | c | c | c |}
		\hline
	\multicolumn{2}{| c |}{$\beta_{1}$}&\multicolumn{2}{ c |}{$\beta_{2}$}\\\hline
		LLG & Thiele &  LLG  &  Thiele \\ \hline
		$-17.1^\circ$  & $-16.8^\circ$ & $57.3^\circ$ & $59.73^\circ$  \\ \hline
	\end{tabular}
\end{table}



The homotopy concept used for constructing hybrid skyrmion tubes can be extended further for solitons localized in all three dimensions.
For that one can consider the homotopy transformation of any 2D skyrmion with $Q=0$ into the FM phase.
Here we examine such transformation on the example of chiral droplet shown in Fig.~\ref{Fig3}.
The MEP in Fig.~\ref{Fig3} is quite similar to that shown in Fig.~\ref{Fig1}.
Two stable states representing the isolated soliton and the saturated FM state are separated by the saddle point.
%
In this transformation, the collapse of the droplet goes by its shrinking, which has been previously discussed in Refs.~\cite{Muckel2021,Kuchkin_2021_v2}.

\begin{figure*}
\includegraphics[width=17.3cm]{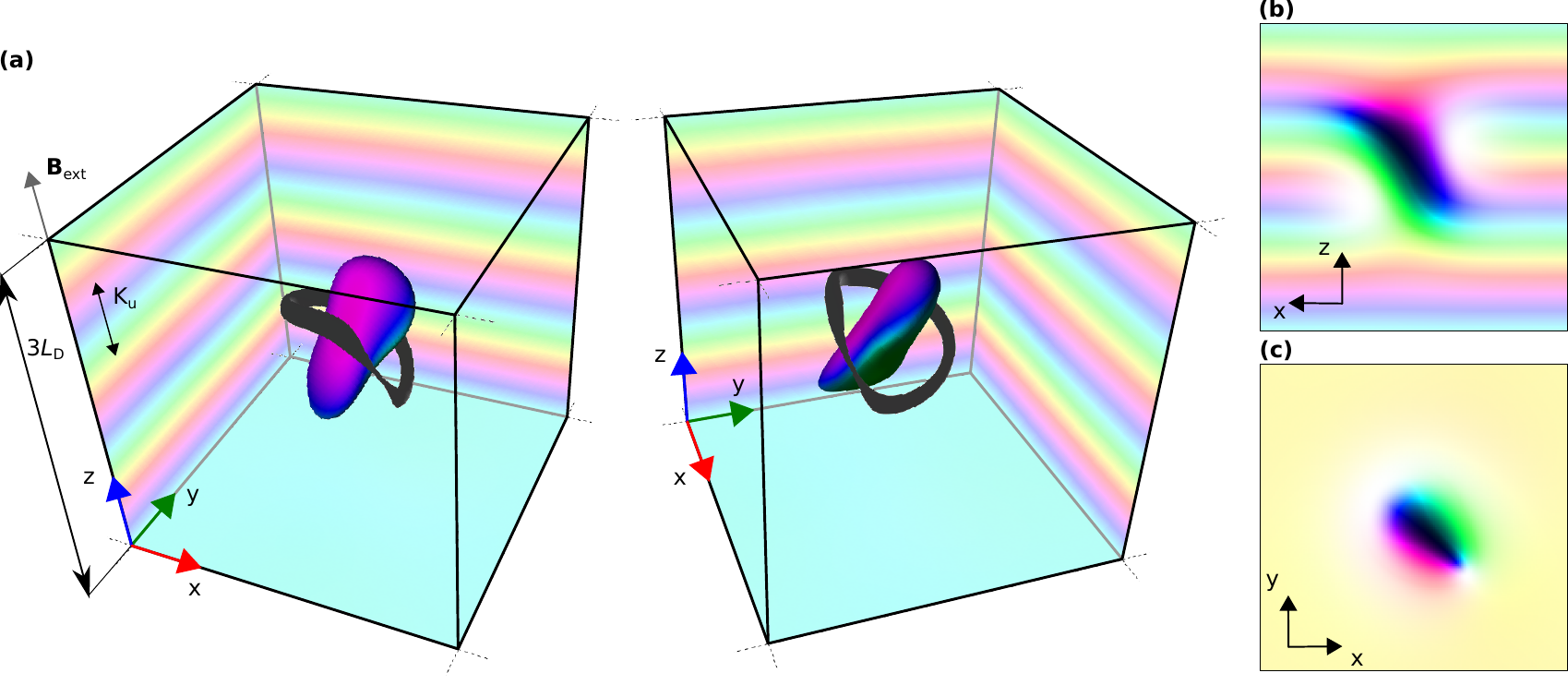}%
\caption{\textbf{a} shows isosurfaces $n_\mathrm{z}=0$ and $n_\mathrm{z}=0.98$ (black) for 3D chiral droplet stabilized at $h=0.34$, $u=0.26$ in the box of size $L_\mathrm{x}=L_\mathrm{y}=L_\mathrm{z}=3L_\mathrm{D}$, $L_\mathrm{D}=64a$ with periodic boundary conditions in all directions.
\textbf{b}, \textbf{d} show the sections of the droplet by planes $y=0$ and $z=0$, respectively.\label{Fig4}}
\end{figure*}

By mapping the obtained images of the corresponding homotopy path onto the spatial $z$-axis, we constructed the initial state of the 3D soliton.
However, it turned out that such an initial guess does not lead to a stable configuration in the FM phase but instead becomes stable in the cone phase vacuum only.
The isosurfaces and cross-sections of the stable configuration of the 3D chiral droplet obtained by numerical energy minimization are shown in Fig.~\ref{Fig4}.
As seen in (\textbf{a}) the red color in the surface is missing, which means that by projecting the vector field of the 3D chiral droplet onto $S_{2}$ sphere, one can not cover it entirely.
The latter means that both topological invariant ~\eqref{Q} and Hopf invariant are zero in this case. 
The cross-sections in (\textbf{b}) and (\textbf{c}) show well-localization of the soliton.
We estimated the characteristic size of the 3D chiral droplet equal to $\sim1 L_\mathrm{D}$  in all spatial directions.

\begin{figure}
\includegraphics[width=8.2cm]{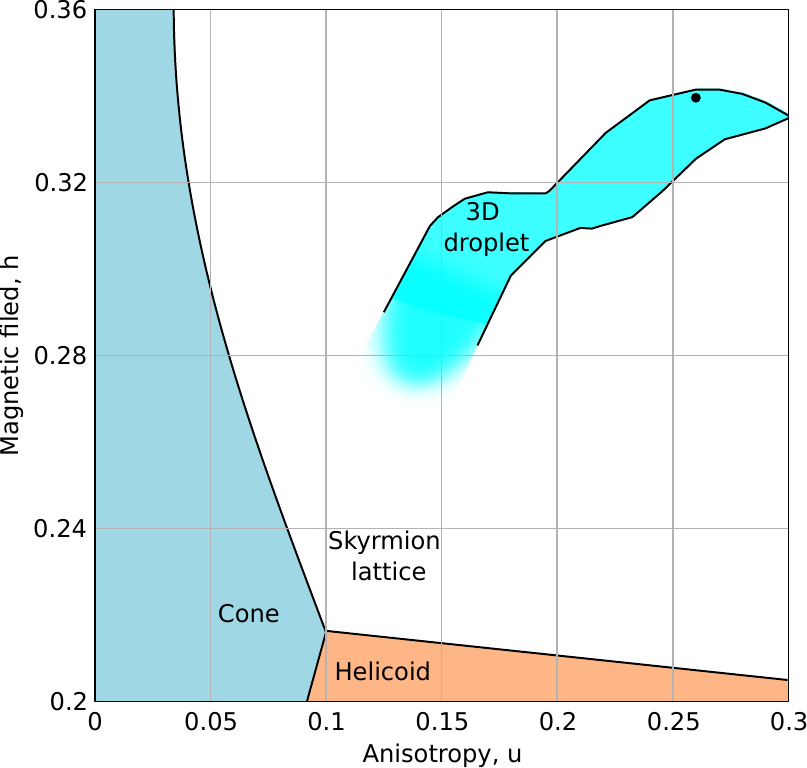}%
\caption{\label{Fig5} The stability range of the 3D chiral droplet (blue) is shown. The phase transition lines are taken from Ref.~\cite{Wilson}.   }
\end{figure}

The stability range for the 3D chiral droplet in terms of the magnetic field, $h$, and anisotropy, $u$, is shown in Fig.~\ref{Fig5}.
Note, the phase transition lines between the phases are reproduced from Ref.~\cite{Wilson}.
As follows from the diagram, the 3D chiral droplet stability region is in the range where the skyrmion lattice is the lowest energy state, while the cone phase is a meta-stable state.
The blurred edges at $h\lesssim0.28$ and $u\lesssim0.15$ denote that we did not succeed to identify the stability range below these values reliably. 
At these values of magnetic field and anisotropy, the helicoid competes in energy with the cone phase.
A more precise estimation of the lower bound for the 3D chiral droplet stability requires much larger sizes of the simulated domain above the limit of our computational resources.

The 3D chiral droplet motion can be induced by applying different external stimuli.
First, let us consider the motion of the 3D chiral droplet induced by the external magnetic field slightly tilted and rotating about the z-axis, $h = 0.34(\sin\theta_\mathrm{h}\cos2\pi\nu t,\sin\theta_\mathrm{h}\sin2\pi\nu t,\cos\theta_\mathrm{h})$. 
We performed the micromagnetic simulations assuming the tilt angle $\theta_\mathrm{h}=0.1$ and the small frequency of rotation $\nu=5$ MHz which is close to a quasi-stationary process. 
Supplementary Movie 4 demonstrates such motion of the 3D chiral droplet consisting of two types of motion -- the translation along the $z$-axis and the rotation about the $z$-axis.
Remarkably, similar dynamics has been reported for heliknoton~\cite{Smalykh_2020} and seems to be a common property of all 3D localized solitons hosted by the conical phase.
The motion of the 3D chiral droplet free of its rotation can be induced by the cone phase rotation at the static external field.
To demonstrate this, we performed the simulations where we manually rotated a small number of spins located far from the droplet.
As illustrated in Supplementary Movie 5, the translation motion of the droplet along the $z$-axis, in this case, occurs without its rotation.
%


In the case of Zhang-Li torque, the gyro vector of the 3D droplet equals zero and the Thiele equation \eqref{Thiele_eq} provides a trivial solution, $\mathbf{v}=\xi\mathbf{I}/\alpha$.
On the other hand, due to the presence of the cone phase, one has to choose $\mathbf{I}\perp \mathbf{q}$ to suppress the excitation of the cone phase itself.
The analysis of the cone phase dynamics induced by $\mathbf{I}\parallel \mathbf{q}$, is provided in Appendix~\ref{A_cone_rot}.

One has to note that contrary to the Thiele equation, which predicts that the 3D chiral droplet should move without deflection, $\mathbf{v}\parallel-\mathbf{I}$, in numerical simulations, we still observe a slight deflection in the soliton motion.
Since, in the case of the hybrid skyrmion tube, we got pretty well agreement between numerical simulations and the Thiele approach, we attribute this effect to the numerical artifact caused by the excitation of the cone phase, which in turn prevents reaching the regime of a steady motion.
Supplementary Movie 6 shows the dynamics of the droplet caused by the in-plane electric current $\mathbf{I}\parallel \mathbf{e}_\mathrm{x}$. To prevent the excitation of the cone phase, in this simulation, we pinned the spins on the bottom plane of the simulated domain.

\section{\label{sec:conclusions}Conclusions}

In this work, we have discussed new types of 3D magnetic solitons stabilized in chiral magnets -- hybrid skyrmion tube and 3D chiral droplet.
We show that the magnetic textures of these solutions can be explained in terms of homotopy transitions between various 2D skyrmions.

We have studied the static and dynamic properties of new solitons.
The parameters of the external magnetic field and magnetic anisotropy at which the presented solutions remain stabilized are different and characterized by the various states of the vacuum.
hybrid skyrmion tubes can be stable in the FM state and cone phase, while the 3D chiral droplet is stable only in the conical phase surroundings.
It is shown that the dynamics of the hybrid skyrmion tube can be induced by the electric current modeled by the Zhang-Li torque in the LLG equation.
The electric current applied along the tube causes the motion of the knot on the skyrmion tube in the direction opposite to the current.
The electric current applied perpendicular to the skyrmion tube leads to more complicated dynamics, which can be described in terms of two skyrmion Hall angles.
The results of LLG simulations agree well with the analysis based on the Thiele equation.

For the case of the 3D chiral droplet, we show that its motion besides the electric current can be induced by the rotating external magnetic field.
Since the cone phase is easily excited when the electric current is not perpendicular to the cone wave vector, the analysis based on the Thiele approach is difficult in this case.
The analytic solution for the dynamics of the cone phase is provided.
Supplementary Movies illustrate numerical simulations for the motion of the 3D solitons. 

\section*{Supplementary Material}
See supplementary material for additional movies which illustrate the dynamics of 3D solitons discussed in the main text. 
Supplementary Movie 1 illustrates the escape of the knot on the hybrid skyrmion tube through the free boundary of the plate.
Supplementary Movie 2 and 3 show the dynamics of the hybrid skyrmion tube induced by current $\mathbf{I}\parallel \mathbf{e}_\mathrm{z}$ and $\mathbf{I}\parallel \mathbf{e}_\mathrm{x}$, respectively.
Supplementary Movie 4 illustrates the dynamics of the 3D chiral droplet induced by the small precession of the external magnetic field about the $z$-axis with low frequency.
Supplementary Movie 5 shows the dynamics of the 3D chiral droplet under the cone spiral rotation induced by a weak electric current $\mathbf{I}\parallel \mathbf{e}_\mathrm{z}$ applied in the bounded volume far from the droplet.
Supplementary Movie 6 shows the dynamics of the 3D chiral droplet induced by current $\mathbf{I}\parallel \mathbf{e}_\mathrm{x}$ when the spins in the plane $z=0$ are pinned.

The magnetization in the movies is presented by the isosurfaces $n_\mathrm{z}=0$ and the standard color code used in Mumax for visualization of the unit vector fields. 
The exceptions are the Movies 4-6 where we use red-blue color code for $n_\mathrm{y}$-component of magnetization: $\mathbf{n}=(0,-1,0)$ -- red and $\mathbf{n}=(0,1,0)$ -- blue.

\begin{acknowledgments}
We are acknowledge Filipp Rybakov, Bernd Schroers, and Bruno Barton-Singer for fruitful discussions. We also grateful to Moritz Sallermann for technical support in minimum energy path calculations with Spirit. 
We acknowledge financial support from the Deutsche Forschungsgemeinschaft through SPP 2137 ``Skyrmionics" Grant No. KI 2078/1-1 and the European Research Council under the European Union's Horizon 2020 Research and Innovation Programme (Grant No.~856538 - project ``3D MAGiC''). 
\end{acknowledgments}

\appendix

\section{Details of MEP calculation}\label{A_MEP}

The MEP in Fig.~\ref{Fig1} \textbf{(a)} represents a homotopy path between two states \textbf{(b)} and \textbf{(d)}.
To get the MEP without singularities, one has to make a reasonable assumption for the initial path.
The straightforward and often used approach based on the interpolation between two configurations by simple rotation of the spins is not appropriate for this purpose.
It is easy to check that the topological index $Q$ is often not conserved along the MEP with this approach.
To avoid such a behavior of the GNEB solver, we have constructed the ansatz for the initial path using the energy minimization and drag-and-drop function implemented in Spirit.
In particular, as seen in Fig.~\ref{Fig1}(\textbf{b})-(\textbf{e}) the intermedeate states represents the elongation and twisting of the magnetic texture near the chiral kink.
Starting with the state depicted in Fig.~\ref{Fig1}(\textbf{b}) and using the drag-and-drop option we enforce the chiral kink to elongate and bend to form a state similar to that in (\textbf{c}) and (\textbf{e}). We intentionally stop the relaxation process before the system converges to one of two local minima as in Fig.~\ref{Fig1}(\textbf{b}) or (\textbf{d}). 
Then, we collect such unrelaxed snapshots of the spin texture with the different levels of elongation of the part containing the chiral kink and use them as an initial guess for MEP.

The calculation were perfomed at the following parameters: $L_\mathrm{x}=L_\mathrm{y}=4L_\mathrm{D}$ with $L_\mathrm{D}=64a$.

\section{Cone phase rotational dynamics}\label{A_cone_rot}

In terms of the spherical angles $(\Theta,\Phi)$ the magnetization writes $\mathbf{n}=(\sin\Theta\cos\Phi,\sin\Theta\sin\Phi,\cos\Theta)^\mathrm{T}$.
Assuming that at $t=0$, the magnetization profile is given by the cone phase $\Theta=\Theta_\mathrm{c}$, $\Phi=2\pi z+\phi_{0}$, we can find the analytic solution of the LLG equation with the current $\mathbf{I}=I\mathbf{e}_\mathrm{z}$ turned on at $t>0$.
The uniformity of the magnetic texture in the $xy$-plane leads to the following system of equations for $(\Theta,\Phi)$:
\begin{widetext}
\begin{equation}
	\begin{cases}
		\dfrac{\partial^2 \Theta}{\partial z^2} -\left(\dfrac{\partial \Phi}{\partial z}\right)^{\!2}\!\dfrac{\sin 2\Theta}{2}
		\!+\!2\pi\dfrac{\partial \Phi}{\partial z}\sin2\Theta-4\pi^{2}u\sin2\Theta
		-4\pi^{2}h\sin\Theta+i\xi\dfrac{\partial \Theta}{\partial z}-\alpha\dfrac{\partial \Theta}{\partial \tau}\!+\!\left(\dfrac{\partial \Phi}{\partial \tau}-i\dfrac{\partial \Phi}{\partial z}\right)\sin\Theta=0, \\
		-2\dfrac{\partial \Theta}{\partial z}\dfrac{\partial \Phi}{\partial z}\cos\Theta-\dfrac{\partial^2 \Phi}{\partial z^2}\sin\Theta+4\pi\dfrac{\partial \Theta}{\partial z}\cos\Theta
		-i\dfrac{\partial \Theta}{\partial z}+\dfrac{\partial \Theta}{\partial \tau}+\left(\alpha\dfrac{\partial \Phi}{\partial \tau} -i\xi\dfrac{\partial \Phi}{\partial z}\right)\sin\Theta=0.
	\end{cases}
\label{LLG_cone_ZL}
\end{equation}
\end{widetext}
%
where $\tau=\gamma B_\mathrm{D}t/4\pi^{2}$ is  the dimensionless time and $i=4\pi^{2}I/\gamma L_\mathrm{D}B_\mathrm{D}$ is the parameter of electric current.
Although the system \eqref{LLG_cone_ZL} represents two coupled non-linear partial differential equations, its solution can be written for the case when $\Theta=\Theta(\tau)$. 

\begin{figure*}[ht!]
\includegraphics[width=17.1cm]{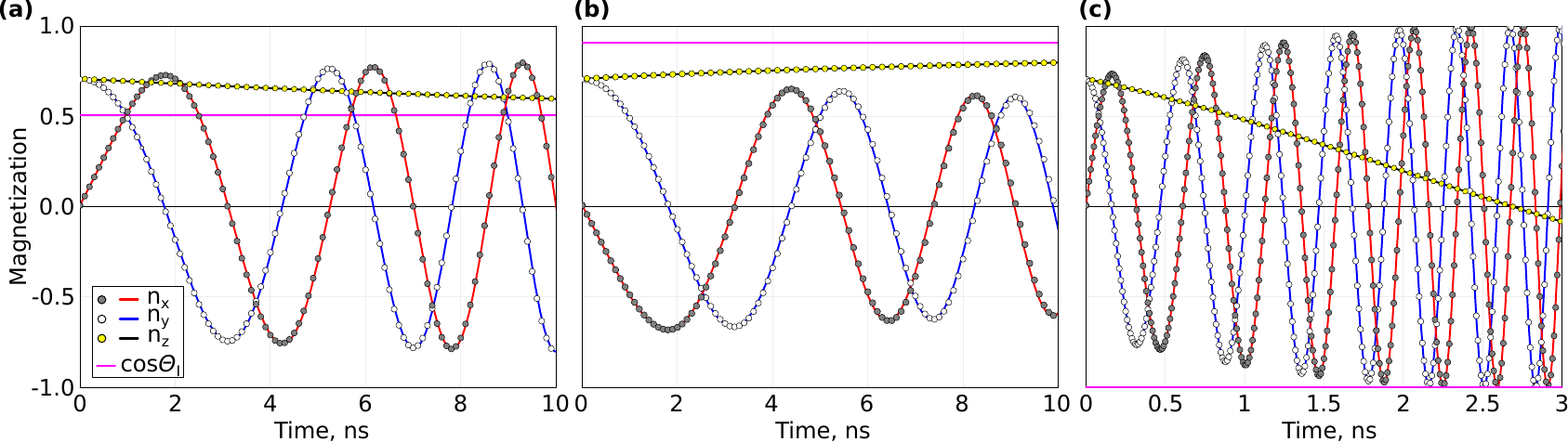}%
\caption{\label{fig:5} \textbf{a}-\textbf{c} show the functional dependencies of magnetization components on time at $z=0$ for current values $i_{1}=-0.15$, $i_{2}=0.15$ and $i_{3}=2$.
Solid lines are given by analytic solutions \eqref{Theta_ZL_sol}, \eqref{Psi_ZL_sol} while points are obtained in numerical simulations with Mumax.
Magenta solid line corresponds to the limit value for $n_\mathrm{z}$ as follows from \eqref{Theta_i}.
}
\end{figure*}

We found that strong currents can lead to the transition from the cone phase to the FM state.
The critical current at which this happens is given by:
\begin{equation}
	i_\mathrm{c}^{\pm}=\dfrac{2\pi\alpha(1-2u)}{\xi-\alpha}(\cos\Theta_\mathrm{c}\pm1),
\label{Ic}
\end{equation}
The formula \eqref{Ic} has sense only at $\xi\neq\alpha$.
In the case $\xi=\alpha$, the cone phase can not be excited by the current.
Below the critical regime, i.e. $i\in(i_\mathrm{c}^{-},i_\mathrm{c}^{+})$, the angle $\Theta$ monotonically changes from $\Theta_\mathrm{c}$ -- the equilibrium cone angle without electric current to $\Theta_I$ -- the equilibrium cone angle in dynamical steady state in the presence of the current:
\begin{equation}
    \cos\Theta_I = \cos\Theta_\mathrm{c}-\dfrac{i(\xi-\alpha)}{2\pi \alpha(1-2u)}.
\label{Theta_i}
\end{equation}
For $i\notin(i_\mathrm{c}^{-},i_\mathrm{c}^{+})$ the angle $\Theta_I$ equals to $0$ or $\pi$ depending on the sign of the current, $i$.
At $i\in(i_\mathrm{c}^{-},i_\mathrm{c}^{+})$, the solution of \eqref{LLG_cone_ZL} for $\Theta$ can be written as: 
\begin{widetext}
\begin{equation}
    \dfrac{8\pi^{2}\alpha \tau}{1+\alpha^2}=
    \begin{cases}
    \dfrac{1}{a_{1}}\dfrac{1}{\cos\Theta_\mathrm{c}\mp1}-\dfrac{1}{a_{1}}\dfrac{1}{\cos\Theta\mp1}+\dfrac{1}{2a_{1}}\ln\left(\dfrac{1+\zeta}{1+\cos\Theta_\mathrm{c}}\dfrac{1-\cos\Theta_\mathrm{c}}{1-\cos\Theta}\right), a_{2}=\pm a_{1}, a_{1}\neq0,\\
   \dfrac{1}{a_{1}^2\!-\!a_{2}^2}\ln\left(\!\left(\!\dfrac{1\!+\!\cos\Theta}{1\!+\!\cos\Theta_\mathrm{c}}\right)^{a_{1}-a_{2}}\!\left(\dfrac{1\!-\!\cos\Theta_\mathrm{c}}{1\!-\!\cos\Theta}\right)^{a_{1}+a_{2}}\!\left(\dfrac{a_{1}\!-\!a_{2}\cos\Theta}{a_{1}\!-\!a_{2}\cos\Theta_\mathrm{c}}\right)^{2a_{2}}\right),|a_{2}|\neq|a_{1}|,
    \end{cases}
    \label{Theta_ZL_sol}
\end{equation}
\end{widetext}
where $a_{1}=h-i(\xi-\alpha)/2\pi\alpha$, $a_{2}=1-2 u$.
The solution for $\Phi$ function is given by:
\begin{equation}
    \Phi(\tau,z)=2\pi\left(z+ \dfrac{i\xi}{\alpha}\tau\right) + \phi_{0}+ \dfrac{1}{\alpha}\ln\dfrac{\tan\left(\Theta_\mathrm{c}/2\right)}{\tan\left(\Theta/2\right)}.
\label{Psi_ZL_sol}
\end{equation}
As follows from \eqref{Theta_ZL_sol}, at $\tau\rightarrow\infty$, one has $\cos\Theta=\cos\Theta_{I} = a_{1}/a_{2}$. In this case, Eq.~\eqref{Psi_ZL_sol} describes the rotation of the cone phase with constant velocity $I\xi/\alpha$ which agrees with the Thiele prediction for topologically trivial configurations.
Noticeably, the dynamics described by the LLG equation reaches the dynamical steady state exponentially fast.
%
Note, the solutions \eqref{Theta_ZL_sol}, \eqref{Psi_ZL_sol} remain valid not only for the bulk systems but also for the films with free boundaries along $z$-axis.
The latter remain valid for any $h$ and $u$ where the surface modulations~\cite{Rybakov2015} do not perturb the cone phase.
This follows from the fact that found solutions satisfy the boundary conditions: 
$\partial_z\Theta=0$, $\partial_z\Phi=2\pi$ on the free surface, $z=\mathrm{const}$, for any $\tau$.

To verify the found solutions, we compare them with the results of LLG simulations performed for different values of the current but fixed values of  $h=0.34$, $u=0.26$, $\alpha=0.01$, and $\xi=0.05$. 
The critical current values \eqref{Ic} for these parameters are: $i_\mathrm{c}^{-}\approx-0.22$, $i_\mathrm{c}^{+}\approx 1.29$.
We have performed simulations for two current values within the critical range: $i_{1}=-0.15$, $i_{2}=0.15$, and for one out of this range $i_{3}=2$.
The results are provided in Fig.~\ref{fig:5}.
For all three cases we see a good agreement between numerical and analytical solutions at least for the chosen simulation time.
At longer times, $z$-component of the magnetization tends to the limit value accordingly to \eqref{Theta_i}.
For the cases shown in (\textbf{a}) and (\textbf{b}) these limit values are about $0.51$ and $0.91$, respectively.
For the case shown in (\textbf{c}) corresponding to the current above the critical value, the magnetization tends to $-1$.
As one can deduce, applying the current $\mathbf{I}\parallel\mathbf{q}$ allows manipulating the cone phase angle and its dynamics in a controllable way.

The found analytical solutions besides the pure academic interests can be used also for testing the accuracy of numerical schemes for solving the LLG equation.
In particular, the solutions of \eqref{LLG_cone_ZL} in a special case of zero damping, $\alpha=0$, can be written in a more compact form:
\begin{eqnarray}
    & \Theta=2\arctan\left(\tan\dfrac{\Theta_\mathrm{c}}{2}e^{2\pi i\xi \tau}\right),\nonumber\\
    & \Phi=2\pi(z+i\tau)+\phi_{0}-4\pi^{2}(1-2u - h)\tau\nonumber\\
    & +\dfrac{2\pi}{i\xi k}\ln\dfrac{1+\tan^{2}\dfrac{\Theta_\mathrm{c}}{2}e^{2\pi i\xi \tau}}{1+\tan^{2}\dfrac{\Theta_\mathrm{c}}{2}}.
    \label{sol_alpha=0}
\end{eqnarray}
The stability of LLG solvers typically requires the presence of non-zero damping.
Therefore, the  solutions \eqref{sol_alpha=0} can be useful for testing new LLG solvers which are free of this limitation.
Both cases $\alpha=0$ and $\alpha\neq0$ can be generalized for the case of time-dependent currents $I=I(t)$ easily.
The detailed analysis of this case is out of the scope of the present study and will be provided elsewhere.

\bibliography{aipsamp}

\end{document}